\documentclass[11pt,a4paper]{article}
\usepackage{amsmath,amssymb}
\usepackage{epsfig,graphicx}

\usepackage{indentfirst}  %  Indentation for first paragraph of each section!

\begin{document}

\title{\bf Primordial black hole formation from non-Gaussian \\ curvature perturbations}
\author{P.~A.~Klimai$^a$\footnote{{\bf e-mail}: pklimai@gmail.com}, \;
E.~V.~Bugaev$^{a}$\footnote{{\bf e-mail}: bugaev@pcbai10.inr.ruhep.ru}
\\
$^a$ \small{\em Institute for Nuclear Research, Russian Academy of Sciences,} \\
\small{\em 60th October Anniversary Prospect 7a, 117312 Moscow, Russia}
}
\date{}
\maketitle

\begin{abstract}
We consider several early Universe models that allow for production of large curvature
perturbations at small scales. As is well known, such perturbations can lead to
formation of primordial black holes (PBHs). We briefly review the today's
situation with PBH constraints and then focus on two models in which strongly
non-Gaussian curvature perturbations are predicted: the hybrid inflation waterfall
model and the curvaton model. We show that PBH constraints on the values of
curvature perturbation power spectrum amplitude are strongly dependent on the
shape of perturbations and can significantly (by two orders of magnitude) deviate
from the usual Gaussian limit ${\cal P}_\zeta \lesssim 10^{-2}$. We give examples
of PBH mass spectra calculations for both inflationary models.
\end{abstract}

%\tableofcontents

\section{Introduction}

As is well known, in models of slow-roll inflation with one scalar field the curvature
perturbation originates from the vacuum fluctuations during inflationary expansion, and
these fluctuations lead to practically Gaussian classical curvature perturbations
with an almost flat power spectrum. However, it is well known also that both
these features are not generic in the case of inflationary models with two (or more)
scalar fields: such models can easily predict adiabatic perturbations with,
e.g., a ``blue'' spectrum and these perturbations can be non-Gaussian \cite{Linde:1996gt}.

Possibilities for appearing of non-Gaussian fluctuations in inflationary models
with multiple scalar fields had been discussed long ago
\cite{Salopek:1988qh, Salopek:1991jq, Fan:1992wv}. The time
evolution of the curvature perturbation on superhorizon scales (which
is allowed in multiple-field scenarios \cite{Starobinsky:1986fxa}) implies
that, in principle, a rather large non-Gaussian signal can be generated during inflation.
According to the observational data \cite{Komatsu:2010fb}, the primordial curvature
perturbation is Gaussian with an almost scale-independent power spectrum. So far there is
only a weak indication of possible primordial non-Gaussianity [at $(2-3)\sigma$ level] from
the cosmic microwave background (CMB) temperature
information data (see, e.g., \cite{Yadav:2007yy}).
However, non-Gaussianity is expected to become an important probe of both the early and
the late Universe in the coming years \cite{Komatsu:2009kd}.

The second important feature of predictions of two-field models is that these
models can lead to primordial curvature perturbations with blue spectrum (for
scales which are smaller than cosmological ones) and, correspondingly, can predict
the primordial black hole (PBH) production at some time after inflation. In this case, PBHs
become a probe for the non-Gaussianity of cosmological perturbations
\cite{Bullock:1996at, Ivanov:1997ia, PinaAvelino:2005rm, Hidalgo:2007vk}. The results of
PBH searches can be used to constrain the ranges of early Universe model parameters.

There are several types of two-field inflation scenarios in which detectable
non-Gaussianity of the curvature perturbation can be generated: curvaton
models \cite{Mollerach:1989hu, Linde:1996gt, Lyth:2001nq, Moroi:2001ct, Lyth:2006gd},
models with a non-inflaton field causing inhomogeneous reheating \cite{Dvali:2003em, Kofman:2003nx},
curvaton-type models of preheating (see, e.g., \cite{Kohri:2009ac} and references therein),
models of waterfall transition that ends the hybrid
inflation \cite{Felder:2000hj, Asaka:2001ez, Copeland:2002ku, GarciaBellido:2002aj}.
In these two-field models the primordial curvature perturbation $\zeta$ has two components:
a contribution of the inflaton (almost Gaussian) and a contribution of the extra field.
This second component is parameterized by the following way \cite{Boubekeur:2005fj}
\begin{equation}
\zeta_\sigma({\bf x}) = a \sigma({\bf x}) + \sigma^2({\bf x}) - \langle \sigma^2 \rangle.
\end{equation}
If $a=0$, one has a $\chi^2$-model. Obviously, the quadratic term can't dominate in $\zeta$ on
cosmological scales where CMB data are available. It can, however, be important on smaller scales.

In the present work we study the predictions of the PBH production for two particular
two-field models: the hybrid inflation model with tachyonic instability
at the end of inflation and the curvaton model.
The potentially large non-Gaussianity in these models is
connected with the fact that the predicted
magnitude of the curvature perturbation is proportional to a square of the
non-inflaton (waterfall or curvaton) field. The blue spectrum in the hybrid
waterfall model can arise, e.g., through the tachyonic amplification due
to the dynamical symmetry breaking \cite{Felder:2000hj} and, in the curvaton
model, due to, e.g., supergravity effects leading to the large effective mass
of the curvaton \cite{Linde:1996gt}.

The main attention in the present paper is paid to a study of probability distribution
function (PDF) of the curvature perturbation and the shape of the black hole mass
function, with taking into account of the non-Gaussianity. The first general study of
PDF of the curvature perturbation in curvaton model was carried out in \cite{Sasaki:2006kq}.

Primordial curvature perturbation spectrum in hybrid inflation
waterfall model was recently studied in 
\cite{Lyth:2010ch, Fonseca:2010nk, Abolhasani:2010kr, Gong:2010zf, Lyth:2010zq, Abolhasani:2011yp, 
Lyth:2011kj, Bugaev:2011qt, Lyth:2012yp},
PBH production in this model was considered in \cite{Lyth:2011kj, Bugaev:2011qt, Lyth:2012yp, Bugaev:2011wy}.
PBH production in curvaton scenario was studied in recent works
\cite{Kawasaki:2012wr, Firouzjahi:2012iz} (without considering
the non-Gaussian effects). The approximate PBH constraints on the curvature
perturbation power spectrum in the curvaton model were obtained in \cite{Lyth:2012yp}.

The plan of the paper is as follows. In Sec. \ref{sec-constr-beta} we briefly review the available
constraints on PBH abundance that follow from different types of astrophysical and cosmological data.
In Sec. \ref{sec-gaussian} we present the recent limits on the amplitudes of
curvature perturbation power spectrum obtained with an assumption of Gaussianity of
primordial curvature perturbations.
In Sec. \ref{sec-waterfall} the hybrid inflation model and constraints on its parameters, coming from
PBH non-observation, are discussed. In Sec. \ref{sec-curvaton} we discuss the possible
production of PBHs in the curvaton model and the corresponding cosmological constraints that
can be obtained. Our conclusions are given in Sec. \ref{sec-concl}.

\section{Available constraints on PBH abundance}
\label{sec-constr-beta}

There are several sources of information that allow to obtain limits on PBH abundance.
In the region of $M_{BH}$ which is of most interest for us ($10^9 \lesssim M_{BH} \lesssim 10^{38}\;$g)
these limits can be divided in three groups:

(i) constraints on PBHs from big bang nucleosynthesis (due to hadron injections
by PBHs \cite{Zeldovich1977}, photodissociation
of deuterium \cite{Lindley1980} and light nuclei, fragmentations of quarks and gluons
evaporated by PBHs \cite{Kohri:1999ex}), $10^9 \lesssim M_{BH} \lesssim 10^{13}\;$g, and from
influence of PBH evaporations on the CMB anisotropy,
$2.5\times 10^{13} \lesssim M_{BH} \lesssim 2.5 \times 10^{14}\;$g \cite{Carr:2009jm},

(ii) constraints on PBHs from extragalactic photon background \cite{Page:1976wx},
$10^{13} \lesssim M_{BH} \lesssim 10^{17}\;$g,

(iii) constraints on non-evaporating PBHs (gravitational and lensing constraints,
$M_{BH} > 10^{15}\;$g).

Constraints on PBHs from data on extragalactic neutrino
background \cite{BugaevD66, Bugaev:2008gw}, in the region
$10^{11} \lesssim M_{BH} \lesssim 10^{13}\;$g, are somewhat weaker than nucleosynthesis constraints.
In the PBH mass region $M_{BH}\sim 10^{35}\;$g there also exist constraints from
non-observation of induced gravitational waves, which have been studied in
\cite{Saito:2008jc, Bugaev:2010bb}.

%%%%%%%%%%%%%%%%%%%%%%%%%%%%%%%%%%%%%%%%%%%%%%%%%%%%%%%%%%%%%
\begin{figure}[!t]
\center %
\includegraphics[width=0.7\columnwidth, trim = 0 5 0 0 ]{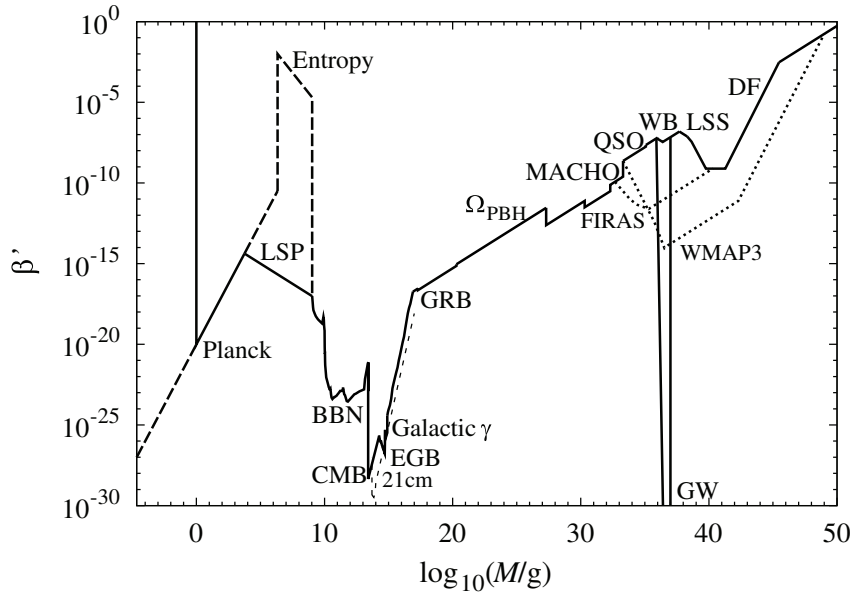}
\caption{ \label{fig-beta}
The available limits on PBH abundance $\beta'(M_{BH})$ from different types
of observations. The figure is taken from \cite{Carr:2009jm}.
}
\end{figure}
%%%%%%%%%%%%%%%%%%%%%%%%%%%%%%%%%%%%%%%%%%%%%%%%%%%%%%%%%%%%%

For a detailed review of these and other PBH constraints, see \cite{Carr:2009jm, Josan:2009qn}.
Traditionally, the constraints are expressed as limits on energy density fraction of
the Universe contained in PBHs at the moment of their formation,
$\beta_{PBH}$, as a function of PBH mass $M_{BH}$ (it is typically assumed that
\begin{equation}
\label{fhMh}
M_{BH} = f_h M_h,
\end{equation}
where $M_h$ is the horizon mass at the moment of PBH formation and $f_h$ is a constant of order of one).
The summary of existing PBH limits,
taken from \cite{Carr:2009jm}, is shown in Fig. \ref{fig-beta}. This Figure gives limits in
terms of $\beta'$, which is simply related to $\beta_{PBH}$ \cite{Carr:2009jm}:
\begin{equation}
\beta'(M_{BH}) = f_h^{1/2} \left( \frac{g_{*}}{106.75} \right)^{-1/4} \beta_{PBH}(M_{BH}),
\end{equation}
where $g_*$ is the effective number of relativistic degrees of freedom.

The available limits on PBH abundance, shown in the figure, can be transformed in constraints on
other cosmological parameters. If, for example, inflationary model predicts large values for the
curvature perturbation power spectrum at some cosmological scale, the model parameters can be constrained
based on the requirement that PBHs are not over-produced.

\section{PBH constraints in the Gaussian case}
\label{sec-gaussian}

Several single-field inflationary models which predict a peaked power spectrum at some scale $k_0$
have been considered in \cite{Saito:2008em, Bugaev:2008bi}. The corresponding constraints on
the amplitudes of the curvature perturbation power
spectrum from the non-observation of PBHs have been obtained in \cite{Bugaev:2010bb}, with
an assumption of Gaussian form of the perturbation distributions.
It had been shown in \cite{Saito:2008em} for the particular case of
Coleman-Weinberg inflationary potential that
non-Gaussianity of perturbations is, really, rather small. Using the assumption of Gaussianity,
the PBH abundances can be calculated with a probability distribution of the form
\begin{equation}
p(\delta_R) = \frac{1}{\sigma_R(M) \sqrt{2\pi}} \exp
\left[-\frac{\delta_R^2} { 2\sigma_R^2(M)} \right],
\end{equation}
where $\sigma_R(M)$ is the mass variance (mean square deviation of density contrast $\delta$ in the
sphere of comoving radius $R$), $M$ is the initial mass of fluctuation.

It is convenient to use some kind of parametrization to model the realistic peaked power
spectrum of finite width. In \cite{Bugaev:2010bb} the distribution of the form
\begin{equation}
\label{PRparam} %
\lg {\cal P}_{\cal R} (k) = B + (\lg {\cal P}_{\cal R}^0 - B)
\exp \Big[-\frac{(\lg k/k_0)^2}{2 \Sigma^2} \Big]
\end{equation}
was used. Here, $B \approx -8.6$, ${\cal P}_{\cal R}^0$ characterizes the height of the peak,
$k_0$ is the position of the maximum and $\Sigma$ is the peak's width.
Note that comoving curvature perturbation ${\cal R}$ practically coincides with
uniform density curvature perturbation $\zeta$ on super-horizon scales.

%%%%%%%%%%%%%%%%%%%%%%%%%%%%%%%%%%%%%%%%%%%%%%%%%%%%%%%%%%%%%
\begin{figure}[!t]
\center %
\includegraphics[width=0.6\columnwidth, trim = 0 5 0 0 ]{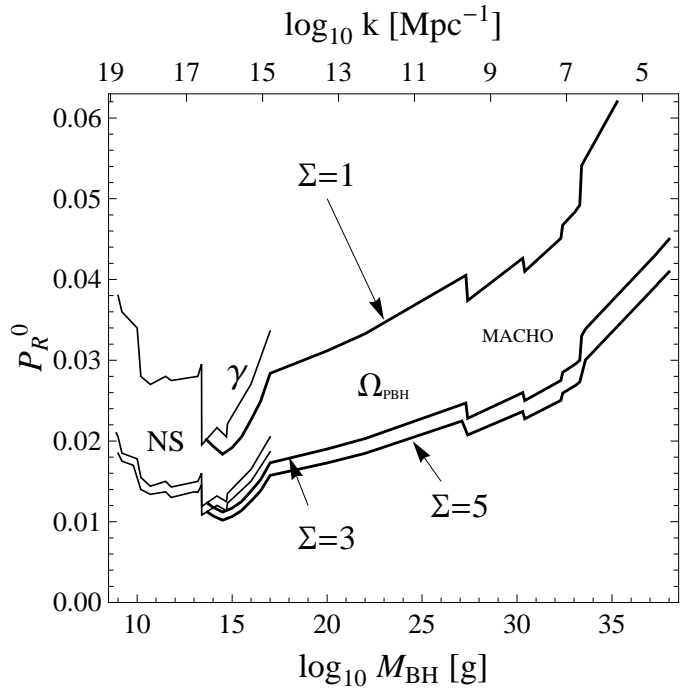}
\caption{ \label{fig-gaussian}
PBH bounds on the maximal value of primordial power spectrum, ${\cal P}_{\cal R}^0$, for the Gaussian case.
Constraints depend on the peak's width $\Sigma$ and its position (which corresponds to PBH
mass $M_{BH}$). The figure is taken from \cite{Bugaev:2010bb}.
} %
\end{figure}
%%%%%%%%%%%%%%%%%%%%%%%%%%%%%%%%%%%%%%%%%%%%%%%%%%%%%%%%%%%%%

Calculating the abundances of PBHs for different sets of model parameters,
and using the available PBH limits \cite{Carr:2009jm}, the constraints on the value of ${\cal P}_{\cal R}^0$
for different values of $\Sigma$ were obtained in \cite{Bugaev:2010bb}.
The corresponding results are shown in Fig. \ref{fig-gaussian}.
It is seen from this Figure that, with the assumption of Gaussianity, the resulting constraints
on the power spectrum amplitude are
of the order of $10^{-2}$ with some scale dependence. Similar results were also
obtained recently in \cite{Josan:2009qn}.

\section{The waterfall transition model}
\label{sec-waterfall}

In this section we consider the hybrid inflation model which describes an evolution of the slowly rolling
inflaton field  $\phi$ and the waterfall field $\chi$, with the
potential \cite{Linde:1991km, Linde:1993cn}
\begin{equation}
\label{pot}
V(\phi, \chi)= \left( M^2 - \frac{\sqrt{\lambda}}{2} \chi^2 \right)^2 + \frac{1}{2}m^2\phi^2
+ \frac{1}{2} \gamma \phi^2 \chi^2 .
\end{equation}
The first term in Eq. (\ref{pot}) is a potential for the waterfall field $\chi$ with the false
vacuum at $\chi=0$ and true vacuum at $\chi_0^2=2 M^2/\sqrt{\lambda} \equiv v^2$. The
effective mass of the waterfall field in the false vacuum state is given by
\begin{equation}
\label{phic-gamma}
m_\chi^2(\phi) = \gamma \left( \phi^2 - \phi_c^2 \right), \qquad
\phi_c^2 \equiv \frac{2M^2\sqrt{\lambda}}{\gamma} .
\end{equation}
At $\phi^2>\phi_c^2$ the false vacuum is stable, while at $\phi^2<\phi_c^2$ the
effective mass-squared of $\chi$ becomes negative, and there is a tachyonic instability
leading to a rapid growth of $\chi$-modes and eventually to an end of the inflationary
expansion \cite{Lyth:2010zq}.
It is convenient to define the following parameters
\begin{equation}
\beta = 2 \sqrt{\lambda} \frac{M^2}{H_c^2} = \frac{|m_\chi^2(0)|}{H_c^2}, \;\;\;\;\;
r = \frac{3}{2} - \sqrt{\frac{9}{4} - \frac{m^2}{H_c^2}},
\end{equation}
where $H_c$ is the Hubble rate during inflation.

It was shown in \cite{Bugaev:2011qt} that the magnitude of the curvature
perturbation can be presented in the form
\begin{equation}
\label{zeta-minusA}
\zeta = \zeta_\chi = - A (\chi^2 - \langle \chi^2 \rangle),
\end{equation}
where $\chi^2$ and $\langle \chi^2 \rangle$ are determined at the time of an
end of the waterfall, $t = t_{end}$, and $A$ is given by the integral
\begin{eqnarray}
\label{Adefined}
A = \int \limits_0^{t_{end}} \frac{H_c dt}{\dot\phi^2(t) + \langle \dot\chi^2(t) \rangle}
\left( \frac{f(t)} {f(t_{end})} \right)^2 \times \qquad \qquad \qquad \qquad \qquad \qquad  \nonumber
\\ \times \frac{1}{2}
 \left[- m_\chi^2(t) +
 \left( \frac{\dot f(t)}{f(t)} \right)^2  -
  \frac{\dot p}{\dot \rho}
  \left( m_\chi^2(t) + \left( \frac{\dot f(t)}{f(t)} \right)^2 \right) \right].
\end{eqnarray}
Here, the function $f(t)$ describes the time evolution of
the waterfall field %%%%%%%%%%%% ------ %%%%%%%%%%%%
(it is proportional to a numerically obtained solution of a linearized field equation for
a Fourier component of a waterfall field perturbation, $\delta\chi_k(t)$,
which is almost independent on $k$).
Energy density $\rho$ and pressure $p$
are a sum of the contributions of $\phi$ and $\chi$ fields.

%%%%%%%%%%%%%%%%%%%%%%%%%%%%%%%%%%%%%%%%%%%%%%%%%%%%%%%%%%%%%
\begin{figure}[!t]
\center %
\includegraphics[width=0.55\columnwidth, trim = 0 5 0 0 ]{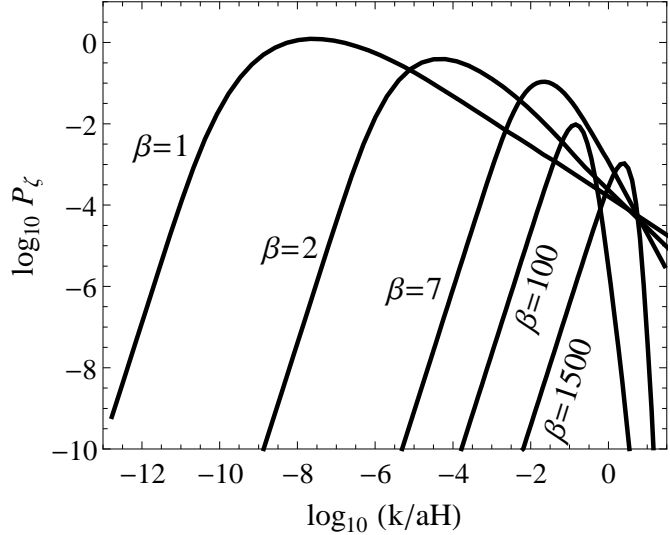}
\caption{ \label{fig-Pzeta}
The numerically calculated spectra ${\cal P}_{\zeta}(k)$ at the moment of the end of the waterfall.
Parameters used for the calculation are:  $r=0.1$, $H_c=10^{11}\;$GeV (all curves);
$\phi_c=3\times 10^{-6} M_P$ (for $\beta=1, 2, 7)$; $\phi_c=0.1 M_P$ (for $\beta=100, 1500)$.
The figure is taken from \cite{Bugaev:2011qt}.
} %
\end{figure}
%%%%%%%%%%%%%%%%%%%%%%%%%%%%%%%%%%%%%%%%%%%%%%%%%%%%%%%%%%%%%

It was shown in \cite{Bugaev:2011qt} that for $\beta \sim 1$ the curvature perturbation
spectrum will reach values of ${\cal P}_\zeta \sim 1$ in a broad interval of other model parameters
(such as $r$, $\gamma$, $H_c$). The peak values, $k_*$, for small $\beta$,
are far beyond horizon, so, the smoothing over the horizon size will
not decrease the peak values of the smoothed spectrum.
The examples of the calculated power spectra for this model are shown in Fig. \ref{fig-Pzeta}.

The relation between curvature perturbation $\zeta$ and the waterfall field value
is given by Eq. (\ref{zeta-minusA}), or, using $\sigma_\chi^2 = \langle \chi^2 \rangle$,
\begin{equation}
\label{zetaAchi2sigchi}
\zeta = - A (\chi^2 - \sigma_\chi^2) = \zeta_{max} - A \chi^2, \qquad \zeta_{max} \equiv A \sigma_\chi^2.
\end{equation}
Here, $A$ and $\sigma_\chi^2$ generally depend on the smoothing scale $R$.
The distribution of $\chi$ is assumed to be Gaussian, i.e.,
\begin{equation}
\label{pchiGauss}
p_\chi(\chi) = \frac{1}{\sigma_\chi \sqrt{2\pi}} \; e^{-\frac{\chi^2}{2 \sigma_\chi^2} }.
\end{equation}

The distribution of $\zeta$ can be easily obtained from (\ref{zetaAchi2sigchi}, \ref{pchiGauss}):
\begin{equation}
\label{pzetaDISTR}
p_\zeta(\zeta) = p_\chi \left| \frac{d\chi}{d\zeta} \right| = \frac{1}{\sqrt{2\pi \zeta_{max}
(\zeta_{max} - \zeta) } } \;
e^{\frac{\zeta - \zeta_{max}}{2\zeta_{max}} }, \qquad \zeta < \zeta_{max},
\end{equation}
which is just a $\chi^2$-distribution with one degree of freedom, with an opposite sign of the argument,
shifted to a value of $\zeta_{max}$. As required, $\langle \zeta \rangle = 0$ and
\begin{equation}
\label{zeta2avg}
\langle \zeta^2 \rangle = \int \limits_{-\infty}^{\zeta_{max}} \zeta^2 p_\zeta(\zeta) d \zeta
= 2 \zeta_{max}^2.
\end{equation}
On the other hand, one has
\begin{equation}
\label{zeta2avgP}
\langle \zeta^2 \rangle = \sigma_\zeta^2 = \int {\cal P}_\zeta(k) W^2(kR) \frac{dk}{k},
\end{equation}
where $W(kR)$ is the Fourier transform of the window function, and we use a Gaussian one,
$W^2(kR) = \exp(-k^2 R^2)$, in this work.

From (\ref{zeta2avg}, \ref{zeta2avgP}) we can write for $\zeta_{max}$ (we now denote the
argument $R$ explicitly):
\begin{equation}
\label{zetamaxexpr}
\zeta_{max}(R) = \left[ \frac{1}{2} \int {\cal P}_\zeta(k) W^2(kR) \frac{dk}{k} \right]^{1/2}.
\end{equation}
Below, we will use the following notation:
$\zeta_{max}(R=0) \equiv \zeta_{max}$. So,
\begin{equation}
\label{zetamax0}
\zeta_{max} = \left[ \frac{1}{2} \int {\cal P}_\zeta(k)\frac{dk}{k} \right]^{1/2}= \frac{1}{\sqrt{2}}
\langle \zeta^2 \rangle^{1/2}.
\end{equation}
It is clear that PBHs can be produced in the early Universe, if $\zeta_{max} > \zeta_c$,
where $\zeta_c$ is the threshold of PBH formation in the radiation-dominated epoch, which
is to be taken from gravitational collapse model. For estimates, in the following we will
use two values: $\zeta_c=0.75$ and $\zeta_c=1$ (corresponding to the PBH formation criterion
in the radiation-dominated epoch: $\delta>\delta_c$, with $\delta_c=1/3$ and $\delta_c=0.45$,
respectively) \cite{Bugaev:2011wy}.

The energy density fraction of the Universe contained in collapsed objects of initial mass
larger than $M$
in Press-Schechter formalism \cite{PS} is given by
\begin{equation}
\label{PSformalism}
\frac{1}{\rho_i} \int\limits_M^{\infty} \tilde M n(\tilde M) d \tilde M = \int\limits_{\zeta_{c}}^{\infty}
p_{\zeta}(\zeta) d\zeta = P(\zeta>\zeta_{c}; R(M), t_i),
\end{equation}
where function $P$ in right-hand side is the probability that in the region of comoving
size $R$ the smoothed value of $\zeta$ will be larger than the PBH
formation threshold value, $n(M)$ is the mass spectrum of the collapsed objects,
and $\rho_i$ is the initial energy density. Here we ignore
the dependence of the curvature perturbation $\zeta$ on time after the end of the waterfall,
assuming it does not change in super-horizon regime, until the perturbations enter horizon
at $k=aH$.

The horizon mass corresponding to the time when fluctuation with initial mass $M$
crosses horizon is (see \cite{Bugaev:2008gw})
\begin{equation}
M_h = M_i^{1/3} M^{2/3},
\end{equation}
where $M_i$ is the horizon mass at the moment $t_i$,
\begin{equation}
\label{Mi}
M_i \approx \frac{4\pi}{3} t_i^3 \rho_i \approx \frac{4\pi}{3} (H_c^{-1})^3 \rho = \frac{4 \pi M_P^2}{H_c}
\end{equation}
(here, we used Friedmann equation, $\rho_i = 3 M_P^2 H_c^2$).
The reheating temperature of the Universe is \cite{Bugaev:2008gw}
\begin{equation}
\label{TRH}
T_{RH} = \left( \frac{90 M_P^2 H_c^2} {\pi^2 g_*} \right)^{1/4}, \qquad g_* \approx 100.
\end{equation}

For simplicity, we will use the approximation that mass of the produced black hole is
proportional to horizon mass [see Eq. (\ref{fhMh})], namely,
\begin{equation}
\label{MBH-Mh}
M_{BH} = f_h M_h = f_h M_i^{1/3} M^{2/3},
\end{equation}
where $f_h \approx (1/3)^{1/2} = {\rm const}$.

Using (\ref{PSformalism}) and (\ref{MBH-Mh}), for the PBH number density (mass spectrum)
one obtains \cite{Bugaev:2011wy}
\begin{equation}
\label{nBH}
n_{BH}(M_{BH}) = \left( \frac{4 \pi}{3} \right)^{-1/3}
\left| \frac{\partial P}{\partial R }\right|
 \frac{f_h \rho_i^{2/3} M_i^{1/3} } {a_i M_{BH}^2}.
\end{equation}

Considering the moment of time for which horizon mass is equal to $M_h$, one can obtain
for the energy density fraction of Universe contained in PBHs:
\begin{equation}
\label{omPBH-beta}
\Omega_{PBH}(M_h) \approx \frac{1}{\rho_i} \left( \frac{M_h}{M_i} \right)^{1/2} \int n_{BH} M_{BH}^2 d \ln M_{BH}.
\end{equation}
It is well known that for an almost monochromatic PBH mass spectrum, $\Omega_{PBH}(M_h)$ coincides with
the traditionally used parameter $\beta_{PBH}$. Although all PBHs do not form
at the same moment of time, it is convenient to use the combination
$M_i^{-1/2} \rho_i^{-1} M_{BH}^{5/2} n_{BH}(M_{BH})$ to have a feeling of how many PBHs
actually form, i.e., to use the estimate following from (\ref{omPBH-beta})
\begin{equation}
\label{beta-estimate}
M_i^{-1/2} \rho_i^{-1} M_{BH}^{5/2} n_{BH}(M_{BH}) \approx \beta_{PBH}.
\end{equation}

%%%%%%%%%%%%%%%%%%%%%%%%%%%%%%%%%%%%%%%%%%%%%%%%%%%%%%%%%%%%%
\begin{figure}[!t]
\center %
\includegraphics[width=0.55\columnwidth, trim = 0 5 0 0 ]{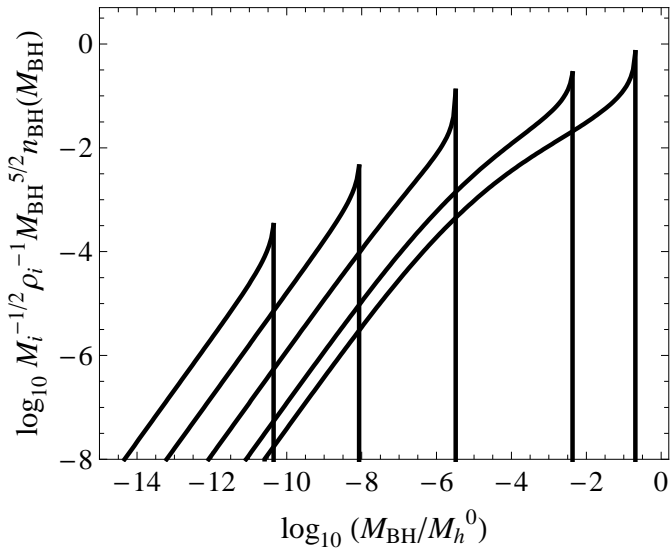}
\caption{ \label{fig-nBH}
The PBH mass spectra for different values of the perturbation spectrum
amplitudes for the hybrid inflation waterfall model. From right to left,
${\cal P}_\zeta^0=1 (\zeta_{max}\approx 1.42), 0.4 (\zeta_{max}\approx 0.9),
0.28 (\zeta_{max}\approx 0.752), 0.27846 (\zeta_{max}\approx 0.750012),
0.2784511 (\zeta_{max}\approx 0.750000068)$.
The position of the peak in ${\cal P}_\zeta(k)$-spectrum is the same for all cases.
For the calculation we used the value $\tilde \Sigma=0.7$, and $\zeta_c=0.75$.
The mass $M_h^0$ corresponds to horizon mass at the moment of time when perturbation with
comoving wave number $k_0$ enters horizon. The figure is taken from \cite{Bugaev:2011wy}.
} %
\end{figure}
%%%%%%%%%%%%%%%%%%%%%%%%%%%%%%%%%%%%%%%%%%%%%%%%%%%%%%%%%%%%%

Fig. \ref{fig-nBH} shows the examples of PBH mass distributions produced in the
waterfall model, calculated assuming
the curvature perturbation power spectrum has the form
\begin{equation}
\label{calP-param}
{\cal P}_\zeta(k) = {\cal P}_\zeta^0 \exp{\left[- \frac{(\lg {k/k_0})^2 } { 2 \tilde \Sigma^2 } \right]},
\end{equation}
where ${\cal P}_\zeta^0$ gives the maximum value approached by the spectrum, $k_0$ is
the comoving wave number corresponding to the position of the maximum and $\tilde \Sigma$ determines
the width of the spectrum (this is just a parametrization of the result of Fig. \ref{fig-Pzeta}).

It is seen from Fig. \ref{fig-nBH} that in the waterfall model PBH abundance severely depends
on the amplitude of the curvature perturbation spectrum in a fine-tuning regime:
once $\zeta_{max}$ is above
$\zeta_c$, PBHs are produced intensively. Demanding
that no excessive amount of PBHs forms in the early
Universe, we can impose the bound on parameters of the inflaton potential.
From the condition $\zeta_{max} < \zeta_c$ one has, for two
fixed values of $\zeta_c$, the following constraints
($r=0.1$; there is a weak dependence on this parameter
but the result is almost independent on $\gamma, H_c$):
\begin{equation}
\label{ec1}
\zeta_c=0.75: \;\; \beta>2.3 , \;\; {\cal P}_\zeta^0 < 0.29 ;
\end{equation}
\begin{equation}
\label{ec2}
\zeta_c=1: \;\; \beta>1.65 , \;\; {\cal P}_\zeta^0 < 0.55 .
\end{equation}
One can see that the limit on the value of ${\cal P}_\zeta$ is significantly weaker
than in the Gaussian case (see Fig. \ref{fig-gaussian}).

\section{A case of the curvaton-type model}
\label{sec-curvaton}

Curvaton is an additional to the inflaton scalar field that can be responsible (partly or fully)
for generation of primordial curvature perturbations
\cite{Mollerach:1989hu, Linde:1996gt, Lyth:2001nq, Moroi:2001ct, Lyth:2006gd}.
This field can also be a source of PBHs, as discussed in 
\cite{Kohri:2007qn, Lyth:2011kj}.

\begin{figure}[!t]
\center %
\includegraphics[width=0.55\columnwidth, trim = 0 5 0 0 ]{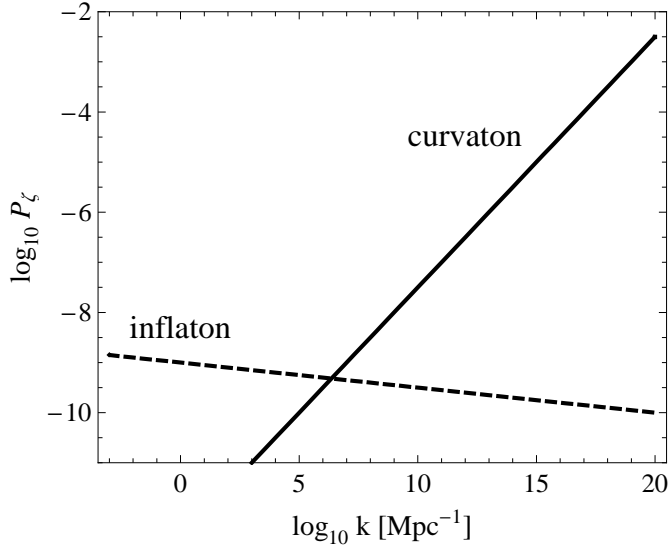}
\caption{ \label{fig-schema}
A sketch that illustrates a relation between curvaton-generated and inflaton-generated
curvature perturbation power spectra for the scenario of PBH production that we consider.
} %
\end{figure}

In this work, we only consider the case of a {\it strong positive tilt}
(the possibility discussed in \cite{Linde:1996gt, Lyth:2006gd})
of the curvaton-generated perturbation
power spectrum. At the same time, it is assumed that inflaton is responsible
for generation of perturbations on cosmological scales (see Fig. \ref{fig-schema} for
an illustration).

%%%%%%%%%%%%%%%%%%%%%%%%%%%%%%%%%%%%%%%%%%%%%%%%%%%%%%%%%%%%%
%%%%%%%%%%%%%%%%%%%%%%%%%%%%%%%%%%%%%%%%%%%%%%%%%%%%%%%%%%%%%

The curvaton field generates cosmological perturbations in two stages
\cite{Linde:1996gt, Lyth:2001nq, Moroi:2001ct, Lyth:2006gd}:

(i) Quantum fluctuations of the curvaton during inflation (at time of
horizon exit) become classical, super-horizon perturbations.

(ii) In the radiation-dominated stage, the curvaton starts to oscillate (this happens
at the time when Hubble parameter becomes of order of curvaton's effective mass, $H\sim m$).
The Universe at this stage becomes a mixture of radiation and matter
(the curvaton behaves as a non-relativistic matter in this regime). The pressure
perturbation of this mixture is non-adiabatic and the curvature perturbation is thus
generated. One obtains, approximately, the expression (see, e.g., \cite{Enqvist:2005pg})
%\begin{equation}
%\zeta \sim r \delta,
%\label{zeta-r-delta}
%\end{equation}
\begin{equation}
\zeta(t, {\bf x}) = \frac{r \sigma_{\rm osc}'} {2 \sigma_{\rm osc}} \delta\sigma_* +
 \frac{r}{4} \left( \frac{\sigma_{\rm osc}' }{\sigma_{\rm osc}} \right)^2 \delta\sigma_*^2,
\label{zeta-r-delta}
\end{equation}
where $r$ is the density parameter, $r= 4\rho_\sigma/(4\rho_r + 3 \rho_\sigma)$
($\rho_r$ is the energy density of radiation after inflation), $\sigma_{\rm osc}$ is
the value of the curvaton field at the onset of oscillations. The initial value
for the curvaton field, $\delta \sigma_*$, is set by inflation. The derivative in
Eq. (\ref{zeta-r-delta}) is taken with respect to the field value during inflation, $\sigma_*$.
The term containing the second derivative, $\sigma_{\rm osc}''$, is neglected.
It is assumed that $r \ll 1$.

Assuming {\it zero average value} for
the curvaton field (i.e., working with the maximal box \cite{Lyth:2006gd}),
we keep in Eq. (\ref{zeta-r-delta}) only the second term,
\begin{equation}
%\delta = \frac{(\delta\sigma)^2}{\langle (\delta\sigma)^2 \rangle}.
\zeta(t, {\bf x}) =
 \frac{r}{4} \left( \frac{\sigma_{\rm osc}' }{\sigma_{\rm osc}} \right)^2 \delta\sigma_*^2.
\label{delta-curv}
\end{equation}
The fluctuations are strongly non-Gaussian which is not forbidden on small scales. Note that the sign
of the perturbation in Eq. (\ref{delta-curv}) is different from the one in Eq. (\ref{zeta-minusA}).

The curvaton-generated curvature perturbation spectrum can be written \cite{Lyth:2006gd} as
(using the Bunch-Davies probability distribution for the perturbations of the
curvaton field \cite{BD, St1982})
\begin{equation}
{\cal P}_{\zeta_\sigma}^{1/2} \sim \frac{2}{3} \Omega_\sigma t_\sigma^{1/2}
\left( \frac{k}{k_e} \right)^{t_\sigma},
\label{Pzeta-curvaton}
\end{equation}
where $\Omega_\sigma = \bar \rho_\sigma / \rho$
is the relative curvaton energy density at the time of its decay (it depends on the
curvaton decay rate) and
\begin{equation}
t_\sigma \cong \frac{2 m_*^2}{3 H_*^2}.
\end{equation}
Here, $m_*$ is the curvaton effective mass during inflation and $H_*$ is the Hubble parameter.

It is rather natural (see \cite{Linde:1996gt} and, e.g., \cite{Demozzi:2010aj},
which considers the
models of chaotic inflation in supergravity) that $t_\sigma \sim 2/3$ which
corresponds to a blue perturbation spectrum with the spectral index
\begin{equation}
n = 1 + 2 t_\sigma \sim 7/3
\end{equation}
(such a situation is shown in Fig. \ref{fig-schema}). For the following, we parameterize the
spectrum (\ref{Pzeta-curvaton}) in a simple form
\begin{equation}
{\cal P}_\zeta = {\cal P}_\zeta ^0 \left( \frac{k}{k_e} \right)^{n-1}, \;\; k< k_e,
\label{Pzeta-curvaton-par}
\end{equation}
and will treat ${\cal P}_\zeta ^0$, $n$, and $k_e$ as free parameters. Note that $k_e$
does not, in general, coincide with the comoving wave number corresponding to the end of inflation. Rather,
it is the scale entering horizon at the time when $\zeta$ is created \cite{Lyth:2006gd}.

\begin{figure}[!t]
\center %
\includegraphics[width=0.55\columnwidth, trim = 0 5 0 0 ]{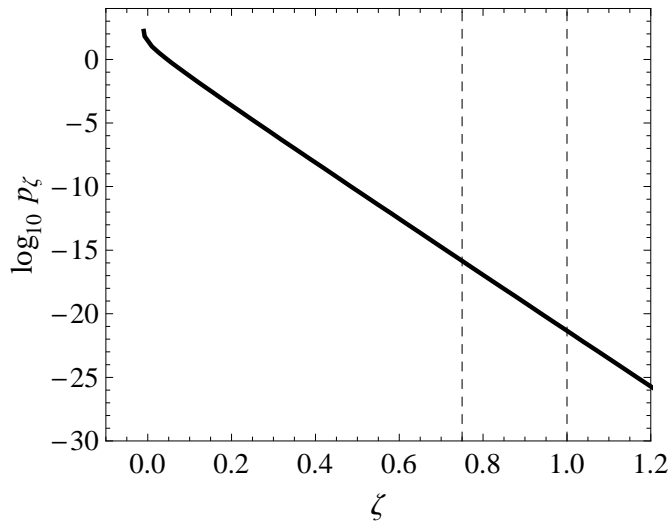}
\caption{ \label{fig-p-zeta-cur}
The form of the distribution (\ref{pzeta-curvaton}) for $\sigma_\zeta^2 = 2\times 10^{-4}$.
Dashed lines show the considered values of $\zeta_c$ ($0.75$ and $1$, see Sec. \ref{sec-waterfall}).
} %
\end{figure}

Using (\ref{delta-curv}) and subtracting the average value, we can write for
the curvature perturbation the formula
\begin{equation}
\zeta = A (\delta \sigma^2 - \langle \delta \sigma^2\rangle),
\end{equation}
and, in analogy with the case of Eq. (\ref{pzetaDISTR}), for the distribution of $\zeta$ one obtains
\begin{equation}
p_\zeta (\zeta) = \frac{1}{ \sqrt{2\pi \zeta_{min} (\zeta_{min} - \zeta) } }
 \; e^{\frac{\zeta - \zeta_{min}} {2 \zeta_{min}} }, \;\;\; {\zeta \ge \zeta_{min}},
 \label{pzeta-curvaton}
\end{equation}
where
\begin{equation}
\zeta_{min} = -A \langle \delta \sigma^2\rangle < 0.
\end{equation}
Note also that
\begin{equation}
\sigma_\zeta^2 = 2 \zeta_{min}^2 = \int{\cal P}_\zeta(k) \frac{dk}{k}.
\end{equation}
The form of the distribution (\ref{pzeta-curvaton}) for $\sigma_\zeta^2 = 2\times 10^{-4}$ is
shown in Fig. \ref{fig-p-zeta-cur}. It is seen that in this particular case the probability to
reach $\zeta \sim \zeta_c \sim 1$ is $\sim 10^{-20}$, i.e., of the same order as
PBH constraints on $\beta'$ of Fig. \ref{fig-beta}. Thus, the value of ${\cal P}_\zeta(k)\sim 10^{-4}$
is already enough to produce an observable amount of PBHs in this model (this is in
agreement with the estimates of \cite{Lyth:2011kj, Lyth:2012yp}).

%%%%%%%%%%%%%%%%%%%%%%%%%%%%%%%%%%%%%%%%%%%%%%%%%%%%%%%%%%%%%
%%%%%%%%%%%%%%%%%%%%%%%%%%%%%%%%%%%%%%%%%%%%%%%%%%%%%%%%%%%%%

For a more exact estimate, we can calculate PBH mass distributions that are generated for a
particular set of parameters ($n$, ${\cal P}_\zeta ^0$, etc.) and then compare the resulting
$\beta_{PBH}$ (using Eq. (\ref{beta-estimate})) with the known limits \cite{Carr:2009jm}
(see Fig. \ref{fig-beta}).

%%%%%%%%%%%%%%%%%%%%%%%%%%%%%%%%%%%%%%%%%%%%%%%%%%%%%%%%%%%%%
\begin{figure}[!t]
\center %
\includegraphics[width=0.5\columnwidth, trim = 0 5 0 0 ]{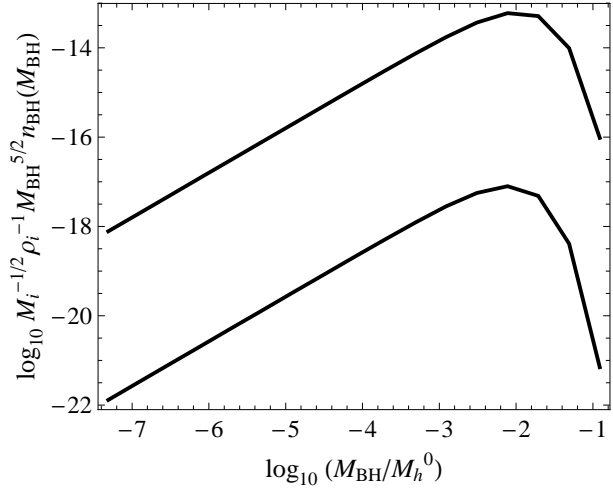}
\caption{ \label{nBH-cur}
Examples of PBH mass spectra calculated for the curvaton model. The parameters used are: $n=2$,
${\cal P}_\zeta ^0 = 4\times 10^{-4}$. For the upper curve, $\zeta_c = 0.75$, for the lower
one, $\zeta_c = 1$. The mass $M_h^0$ corresponds to horizon mass at the moment of time
when perturbation with comoving wave number $k_e$ enters horizon.
} %
\end{figure}
%%%%%%%%%%%%%%%%%%%%%%%%%%%%%%%%%%%%%%%%%%%%%%%%%%%%%%%%%%%%%

The example of PBH mass spectrum calculation is shown in
Fig. \ref{nBH-cur}. The formula (\ref{nBH}) from the
previous Section is used for this calculation (using $p_\zeta$ of Eq. (\ref{pzeta-curvaton})
for the calculation
of the probability $P$).
It is seen from Fig. \ref{nBH-cur} that PBH
abundances depend on particular choice of $\zeta_c$.

\begin{figure}[!b]
\center %
\includegraphics[width=0.6\columnwidth, trim = 0 5 0 0 ]{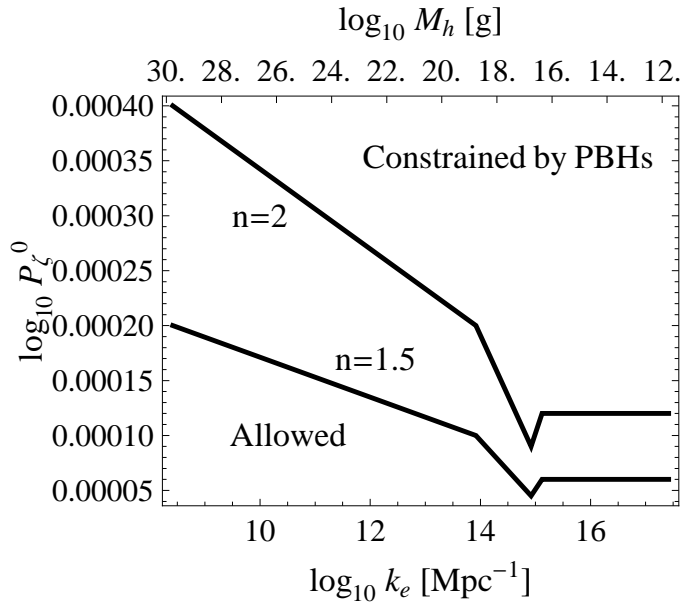}
\caption{ \label{fig-curvaton-limits}
The limits on the maximum value of curvature perturbation power spectrum ${\cal P}_\zeta^0$
from PBH non-observation, for the curvaton model. The forbidden regions are above
of the curves. $M_h$ is the horizon mass at the moment of time when the perturbation with
comoving wave number $k_e$ enters horizon.
The spectrum is parameterized as (\ref{Pzeta-curvaton-par}), $\zeta_c$=0.75.
} %
\end{figure}

The resulting constraints on parameter ${\cal P}_\zeta ^0$
(for two values of the spectral index $n$) from PBH
non-observation, that were obtained in this work, are shown in Fig. \ref{fig-curvaton-limits}.
Thus, we see that the available constraints on the PBH abundance can be transformed into limits
on the power spectrum amplitude and, subsequently, on curvaton model's parameters, such as
$\Omega_\sigma$. For example, comparing Fig. \ref{fig-curvaton-limits} and Eq. (\ref{Pzeta-curvaton}),
one obtains, roughly,
\begin{equation}
\Omega_\sigma  \approx  ({\cal P}_\zeta ^0 )^{1/2}  \lesssim 10^{-2},
\end{equation}
which is already a useful constraint. It is a subject of our further study to get more exact
limits for particular parameter sets of the model.

%%%%%%%%%%%%%%%%%%%%%%%%%%%%%%%%%%%%%%%%%%%%%%%%%%%%%%%%%%%%%
%%%%%%%%%%%%%%%%%%%%%%%%%%%%%%%%%%%%%%%%%%%%%%%%%%%%%%%%%%%%%

\section{Conclusions}
\label{sec-concl}

Primordial black holes can be used to probe perturbations in our Universe at very small scales,
as well as other problems of physics of early stages of the
cosmological evolution. We have considered the PBH
formation from primordial curvature perturbations produced in two cosmological models: in the waterfall
transition in hybrid inflation scenario and in the curvaton model.
Both models which were considered in the paper predict the production of
strongly non-Gaussian perturbations, and non-Gaussianity was taken into account in the calculation
of PBH mass spectrum (in Press-Schechter formalism).

For both models, limits on the values of perturbation power spectrum as well as constraints
on inflation model parameters were obtained. The approximate constraints on the curvature perturbation
spectrum amplitude follow from Eqs. (\ref{ec1}, \ref{ec2}) in the case of hybrid inflation waterfall model and
from Fig. \ref{fig-curvaton-limits} in the case of the curvaton models. One can conclude
that the constraint on ${\cal P}_\zeta$ in the case of hybrid inflation waterfall model is very weak,
${\cal P}_\zeta \lesssim 1$. The corresponding constraint in the curvaton model is much stronger,
${\cal P}_\zeta \lesssim (10^{-4} - 10^{-3})$, depending on the value of the spectral index and
PBH mass. The PBH constraints obtained in this work confirm the estimates given in 
\cite{Lyth:2010zq, Lyth:2012yp}.

\section*{Acknowledgments}

The work was supported by The Ministry of education and science of Russia, project No. 8525.

%\begin{thebibliography}{000} %for 3 digits

\end{document}